\documentstyle[prl,aps,epsf,multicol]{revtex}
\newcounter{saveeqn}

\begin{document}

\setlength{\baselineskip}{12pt}
\title{Adjusting the melting point of a model system via Gibbs-Duhem 
integration: application to a model of Aluminum}

\author{Jess B. Sturgeon and Brian. B. Laird\cite{brian}}
\address{Department of Chemistry and Kansas Institute for
         Theoretical and Computational Science \\
         University of Kansas \\
         Lawrence, Kansas 66045, USA}
\date{\today}
\maketitle
\begin{abstract}

Model interaction potentials for real materials are generally optimized 
with respect to only those experimental properties that are easily 
evaluated as mechanical averages (e.g., elastic constants (at T=0 K), 
static lattice energies and liquid structure).  For such potentials, 
agreement with experiment for the non-mechanical properties, such as 
the melting point, is not guaranteed and such values can deviate 
significantly from experiment.  We present a method for re-parameterizing 
any model interaction potential of a real material to adjust its melting 
temperature to a value that is closer to its experimental melting 
temperature.  This is done without significantly affecting the mechanical 
properties for which the potential was modeled.  This method is an 
application of Gibbs-Duhem integration [D. Kofke, Mol.  Phys.{\bf 78}, 
1331 (1993)].  As a test we apply the method to an embedded atom model 
of aluminum [J. Mei and J.W. Davenport, Phys. Rev. B {\bf 46}, 21 (1992)] 
for which the melting temperature for the thermodynamic limit is $826.4 \pm 
1.3K$ - somewhat below the experimental value of 933K.  After 
re-parameterization, the melting temperature of the modified potential is 
found to be $931.5K \pm 1.5K$. 
\end{abstract}

\ifpreprintsty
\newpage
\else
\begin{multicols}{2}
\fi

\section{Introduction} \label{sec:Introduction}
The ability of a simulation to successfully predict the properties of
real materials is primarily dependent upon the accuracy of the model
interaction potential used.  The construction of model interactions
generally involves the optimization of the parameters of the potential
with respect to {\it mechanical} properties of the material (crystal 
lattice constants, elastic constants, liquid density, etc.) as 
determined from experiment or {\it ab initio} calculations.  Non-mechanical 
properties (i.e. those not obtainable as local averages over coordinates) 
such as phase transition temperatures are difficult to include in such 
optimization procedures and are generally calculated for the optimized 
model {\it a posteriori}, and the agreement of such quantities with 
experiment is not guaranteed.  However, for some applications that 
deal directly with such properties, such as in the study of solid-liquid 
interfaces\cite{Laird98} in which the melting temperature plays an 
obviously important role, it is desirable to develop efficient 
procedures for including such non-mechanical properties in the 
optimization.  In this work, we outline a general procedure for adjusting 
the  potential parameters for a system designed to model a real system 
to improve the agreement of the melting point of that system with the 
experimental value. As an example we present an application to an existing 
embedded atom model of aluminum~\cite{Mei92}. 

Our re-parameterization scheme is based on the powerful Gibbs-Duhem 
integration method developed by Kofke\cite{Kofke93a,Kofke93b}.  In 
this technique, the derivative along the coexistence curve of any 
coexistence condition (such as melting temperature or pressure) 
with respect to any parameter of the potential can be determined 
by an appropriate configurational average at a previously determined 
melting point.  The method is generated by the integration of a 
generalized Gibbs-Duhem equation and the steps are analogous to the 
derivation of the familiar Clapeyron equation for the slope of the 
P-T coexistence curve.  Gibbs-Duhem integration has been shown to 
be quite successful in efficiently determining the coexistence 
conditions for entire classes of potentials. For example, the phase 
diagram for the class of  repulsive inverse power potentials, $ u(r) 
= \epsilon \left(\frac{\sigma}{r}\right)^n$, was determined\cite{Agrawal95} 
by starting with the known hard-sphere ($n = \infty$) coexistence 
and integrating the derivative of coexistence curve with respect 
to the parameter $s \equiv 1/n$.  The method has also proved useful 
in a variety of other applications~\cite{Hagen93,Meijer94,Haag94}.

In the current application, one begins with a model potential, 
parameterized for a real system in the usual way with respect to 
mechanical properties of the real system. The melting temperature 
(or pressure) for the model system is then calculated by 
thermodynamic integration. Once this is done, the derivative of 
the melting temperature (or pressure) with respect to all parameters 
of the system can be determined via separate simulations on the 
coexisting fluid and solid using the Gibbs-Duhem procedure.  
The calculated derivatives allow an assessment of the effect of 
each individual parameter on the melting point. From this 
information an appropriate scheme to adjust the parameters to 
improve the melting point can be devised in such a way that the 
agreement with the other experimental properties is  not 
unacceptably compromised.  The Gibbs-Duhem integration and our 
re-parameterization scheme is outlined in more detail in the next 
section. 

As a test application of this procedure, we examine an embedded 
atom model of aluminum  developed by Mei and Davenport\cite{Mei92}. 
This particular model was chosen for three reasons: First, the 
importance of aluminum as a material makes the development of an 
accurate model potential for simulation purposes desirable. 
Second, the large number of parameters and complicated nature of 
the embedded atom potential increases the need for a systematic, 
as opposed to {\it ad hoc} procedure for adjusting the melting point. 
In addition, the zero-pressure melting point for the Mei-Davenport 
potential has been previously determined\cite{Mei92} to be 800$\pm9$K 
somewhat lower than the experimental melting point of aluminum at 
933K. (Note that, the melting point determined by Mei and Davenport 
was calculated for a 256 particle system - the actual value for 
this potential in the thermodynamic limit is slightly higher at 
826.4$\pm$1.3K.)  We find that, in this specific case, only one 
of the parameters of the potential has any significant effect on 
the potential and that changing this parameter according to the 
Gibbs-Duhem procedure yields a new model with the correct 
experimental melting point with no significant change in the 
quality of the agreement of the quantities with respect to which 
the model was originally optimized. Details of this calculation 
as well as a description of the model can be found in 
Section~\ref{sec:Ourmethod} below.  

\section{Gibbs-Duhem Integration and  Melting Point Optimization} 
\label{sec:Ourmethod}

The technique of Gibbs-Duhem integration has been well described 
previously by Kofke\cite{Kofke93a}, but in the interests of 
completeness and clarity we repeat the basic derivation here.  
Consider a single-component system with an arbitrary interaction 
potential, ${\cal U}(\{{\bf R_i}\},\{X_i\})$, where the ${\bf R_i}$ 
are the atomic coordinates and the $X_i$ are the parameters that 
define the potential - no restriction to pairwise additivity need 
be assumed.  Assume there are two phases $\alpha$ and $\beta$ in 
coexistence at a temperature $T$ and pressure $P$.  On the surface 
of coexistence, the chemical potentials (molar Gibbs free energies) 
of the two phases must be equal.  To quantify how changes in $P$,$T$, 
and $X_i$ will affect the chemical potential one can define a 
generalized Gibbs-Duhem equation,
\begin{equation}
d\mu = -s dT + v dP + \sum_i \lambda_i dX_i \\
\label{dmu}
\end{equation}
where $\mu$ is the chemical potential, $s$ and $v$ are the entropy 
and volume per particle, respectively, and the $\lambda_i$ are 
generalized  thermodynamic variables conjugate to the potential 
parameters, ${X_i}$, defined as
\begin{equation}
N \lambda_i \equiv \left (\frac{\partial G}{\partial X_i} \right 
  )_{T,P,X_{j,j\ne i}}
\label{lambdadef}
\end{equation}

Now as one moves infinitesimally away from the original coexistence 
point $(P,T,\{\lambda_i\})$ to another point $(P + dP, T + dT, 
\{\lambda_i + d\lambda_i \})$ on the surface of coexistence, the 
change in $\mu$ must be identical in both phases.  This condition 
together with Eq.~\ref{dmu} gives
\ifpreprintsty
\begin{equation}
\mu_{\alpha} - \mu_{\beta} = -(s_{\alpha} - s_{\beta}) dT + 
  (v_{\alpha} - v_{\beta}) dP + \sum_i (\lambda_{\alpha,i} - \lambda_{\beta,i})
  dX_i = 0 \; ,\label{cclap} 
\end{equation}
\else
\begin{eqnarray}
\mu_{\alpha} - \mu_{\beta} &=& -(s_{\alpha} - s_{\beta}) dT + 
     (v_{\alpha} - v_{\beta}) dP \nonumber \\
& & \; + \sum_i (\lambda_{\alpha,i} - \lambda_{\beta,i})
  dX_i = 0 \; ,\label{cclap} 
\end{eqnarray}
\fi
where $\mu_{\alpha}$ and $\mu_{\beta}$ are the chemical potentials 
for each of the respective phases.  Assuming constant pressure 
($dP = 0$), since we are interested here in changes in the 
transition temperature, the previous equation can be rearranged to 
give
\begin{equation}
\left (\frac{\partial T}{\partial X_i}\right )_{P,X_{j;j \ne i};\mbox{coex}} 
    = \frac{T (\lambda_{\alpha i} - \lambda_{\beta i})}
    {h_{\alpha} - h_{\beta}}  = \frac{T \Delta \lambda_i}{\Delta h}
\label{gd:deriv}
\end{equation}
where we have also assumed that at coexistence, $\Delta s = \Delta h \; 
/ \; T$, where $\Delta h$ is the latent heat per particle for the phase 
transition.  (Note that, the corresponding equations for the $\left ( 
\frac{\partial P}{\partial \lambda_i} \right )_{T,\lambda_{j;j \ne 
i};\mbox{coex}}$ can be easily obtained by replacing $\Delta h/T$ in 
Eq.~\ref{gd:deriv} with $\Delta v$.)  

The $\lambda_i$ can be related to mechanical averages that can be 
easily calculated in a molecular dynamics or Monte Carlo simulation. 
First, the Gibbs free energy is related to the isothermal-isobaric 
distribution, $\Delta(N,P,T)$ , as follows.
\begin{equation}
G = -k_b T \ln{ \Delta (N,P,T) }
\label{GNPT}
\end{equation}
which for a classical system with interaction potential ${\cal U}
(\{{\bf r}_i \})$ is given by
\begin{equation}
\Delta(N,P,T) = \frac{1}{\Lambda^{3N} N!} \int_0^{\infty} dV 
  \int d^N{\bf r} \; \exp{(-\beta \; {\cal U} - \beta PV)} \; ,
\end{equation}
where $V$ is the volume. 
Taking the derivative of Eq.~\ref{GNPT} with respect to the parameter 
$X_i$  gives

\ifpreprintsty
\begin{equation}
\left( \frac{\partial G}{\partial X_i} \right)_{T,P,X_{j;j\ne i}} = -kT 
\left( \frac{\partial \ln \Delta}{\partial X_i} \right)_{T,P,X_{j;j\ne i}} 
= \frac{1}{\Delta} \int_0^{\infty} dV \int d^N{\bf r} \; \left( 
  \frac{\partial {\cal U}} {\partial X_i} \right)_{X_j;j\ne i} \exp{(-\beta 
  {\cal U} - \beta PV)}  \;.
\end{equation}
\else
\begin{eqnarray}
\left( \frac{\partial G}{\partial X_i} \right)_{T,P,X_{j;j\ne i}} &=& -kT 
\left( \frac{\partial \ln \Delta}{\partial X_i} \right)_{T,P,X_{j;j\ne i}} \nonumber \\
&=&\; \frac{1}{\Delta} \int_0^{\infty} dV \int d^N{\bf r} \; \left( 
    \frac{\partial {\cal U}} {\partial X_i} \right)_{X_j;j\ne i} \nonumber \\
& &\; \times \exp{(-\beta {\cal U} - \beta PV)}  \;.
\end{eqnarray}
\fi

Using Eq.~\ref{lambdadef} we have 
\begin{equation}
\lambda_i = \left( \frac{dG}{dX_i} \right)_{T,P,X_{j;j\ne i}}
        = \left< \left (\frac{\partial {\cal U}}{\partial X_i} 
          \right )_{X_{j; j\ne i}}\right>_{N,P,T} 
\label{lamf}
\end{equation}

Using Eq.~\ref{gd:deriv} and Eq.~\ref{lamf} we can calculate the change 
necessary in each of the parameters $\{X_i\}$, to alter the melting 
temperature of our interaction potential by some arbitrary amount.  
As long as the changes are not too large, the calculation can be greatly
 simplified by the assumption that the deviation remains linear.  
\begin{equation}
T_m(\{X_i\}) = T_{m,0}  + \sum_i \left (\frac{\partial T}{\partial X_i} 
\right )_{X_{j,j\ne i},\mbox{coex}} (X_i - X_{i,0})   \; ,
\end{equation}
where $T_{m,0}$ is the melting point of the original model with parameters 
$\{ X_{i,0} \}$.  (If the linear approximation does not hold  - and it 
should always be checked - one could integrate the differential equation 
using an appropriate numerical technique.) From a single simulation, one 
could calculate the necessary change in all the different parameters of 
the potential corresponding to a particular change in the melting temperature.  
One could then, use this information to construct a cost function to 
re-optimize the potential that includes the previously used set 
$\{A_j^{\mbox{exp.}} \}$ of experimentally determined (mechanical) properties 
as well as the experimental melting temperature $T_m^{\mbox{exp.}}$ as 
optimization targets.  For example, a linear least squares procedure could 
be utilized using appropriately chosen set of weight functions $w_i$: 
\ifpreprintsty
\begin{equation}
f(\{X_i\}) = w_T [T_m^{\mbox{exp.}} - T_m (\{X_i\})]^2 + 
\sum_j w_i [A_j^{\mbox{exp.}}- A_j(\{X_i\})]^2  \;.
\end{equation}
\else
\begin{eqnarray}
f(\{X_i\}) &=& w_T [T_m^{\mbox{exp.}} - T_m (\{X_i\})]^2 \nonumber \\
& & \; + \sum_j w_i [A_j^{\mbox{exp.}}- A_j(\{X_i\})]^2  \;.
\end{eqnarray}
\fi
(In the example application discussed in the next section, we found 
that only one of the several parameters of the potential had any 
significant effect on the melting point, which greatly simplified the 
re-parameterization by making it possible to modify the potential 
without having to directly use such a cost function. It is not expected, 
however, that this will be true, in general.)

\section{Results for a model of Aluminum: determining the original 
         melting point} \label{sec:Results1}
As an application of the method we examine an embedded atom 
model~\cite{Johnson88} for aluminum developed by Mei and Davenport~\cite{Mei92}. 
The parameters in their potential were fit in order to optimize the potential 
with respect to a variety of experimental properties such as the cohesive 
energy ($E_c$), lattice constant ($\sqrt{2}r_0$), un-relaxed vacancy-formation 
energy, and elastic constants of the static fcc crystal at zero temperature.  
The total potential energy has the form
\begin{equation}
{\cal U} = \sum_{i} F({\rho_{i}}) + \frac{1}{2} \sum_{i,j \not= i} \phi 
  (r_{\em ij}) \;, \label{meiU}
\end{equation}
where 
\ifpreprintsty
\begin{eqnarray}
F(\rho) &=& -E_c \left[ 1 - \frac{\alpha}{\beta} \ln{\left( \frac{\rho}{\rho_e} 
  \right)} \right] \left( \frac{\rho}{\rho_e} \right )^{\alpha / \beta} 
+ \frac{1}{2} \phi_0  \sum_{m=1}^{3} s_m exp[ - (\sqrt{m} - 1 ) \gamma] 
  \nonumber \\
& & \; \times \left[ 1+(\sqrt{m} -1) \delta - \sqrt{m} \left( 
  \frac{\delta}{\beta} \right ) \ln{ \left( \frac{\rho}{\rho_e} \right)} 
  \right] \left( \frac{\rho}{\rho_e} \right) ^{\sqrt{m} \gamma / \beta} \; ,
\end{eqnarray}
\else
\begin{eqnarray}
F(\rho) &=& -E_c \left[ 1 - \frac{\alpha}{\beta} \ln{\left( \frac{\rho}{\rho_e} 
  \right)} \right] \left( \frac{\rho}{\rho_e} \right )^{\alpha / \beta} \nonumber \\
& & + \frac{1}{2} \phi_0  \sum_{m=1}^{3} s_m exp[ - (\sqrt{m} - 1 ) \gamma] 
  \nonumber \\
& & \; \times \left[ 1+(\sqrt{m} -1) \delta - \sqrt{m} \left( 
  \frac{\delta}{\beta} \right ) \ln{ \left( \frac{\rho}{\rho_e} \right)} 
  \right] \nonumber \\
& & \; \times \left( \frac{\rho}{\rho_e} \right) ^{\sqrt{m} \gamma / \beta} \; ,
\end{eqnarray}
\fi
and
\begin{eqnarray}
{\rho_{i}} &=& \sum_{j (\not= i)} f(r_{ij}) \;, \label{meip} \\
f(r) &=& {\rho_{e}} \sum_{l = 0}^{5} \frac{c_l}{12} \; \left( \frac {r_0}{r} 
  \right)^{l} \;, \label{fexp} \\
\phi(r) &=& - \phi_{0} \left[ 1 + \delta \left( \frac{r}{r_0} -1 \right) \right] 
  exp \left[ - \gamma \left( \frac{r}{r_0} -1 \right) \right] \; ,
\end{eqnarray}
and is used in conjunction with the following cutoff function:
\begin{eqnarray}
q(r) = \left\{ \begin{array} {r@{\quad \quad}l }
1 & r \leq r_n \; , \\
(1-x)^3 (1 + 3x + 6x^2) & r_n < r < r_c \; , \\
0 & r \geq r_c
\end{array} \right.
\end{eqnarray}
\begin{eqnarray}
x = (r-r_n) / (r_c - r_n)
\end{eqnarray}
with a cutoff distance $(r_c)$ between the third and fourth neighbor 
shells of a static fcc crystal.  Using values of $r_n = 1.75 r_0$ 
and $r_c = 1.95 r_0 \;$, the functions $q(r)f(r)$ and $q(r) \phi(r)$ 
go smoothly to zero at $r=r_c$.  The parameters of the potential
\cite{Mei92,Mei91} are given in Table~\ref{table:Pparams}. For this 
embedded atom model, mass is measured in amu, distance in \AA, and 
energy in eV.  The natural simulation time unit is calculated to be 
10.181fs. 

To begin, the melting temperature of the original model (at $P = 0$) 
must be determined.  Mei and Davenport perform a calculation of the 
free energies of the aluminum melt and fcc crystal using thermodynamic 
integration of the Gibbs free energy. For a 256 particle system, they 
obtain a melting temperature of 800$\pm$9K.
\ifpreprintsty
\else
\end{multicols}
\begin{table}
\bf Table~\ref{table:Pparams}:  Parameters for the Aluminum embedded
atom potential developed by Mei and Davenport.  
\vspace{0.5cm}
\begin{tabular}{c@{\hspace{0.35cm}}c@{\hspace{0.35cm}}c@{\hspace{0.35cm}}
  c@{\hspace{0.35cm}}c@{\hspace{0.35cm}}c@{\hspace{0.35cm}}c@{\hspace{0.35cm}}}
\hline \hline
$E_c$ & $\phi_0$ & $r_0$ & $\alpha$ & $\beta$ & $\gamma$ & $\delta$ \\
\hline
$3.39$ & $0.1318$ & $2.8638$ & $4.60$ & $7.10$ & $7.34759$ & $7.35$ \\
$c_0$ & $c_1$ & $c_2$ & $c_3$ & $c_4$ & $c_5$ \\
$0.64085$ & $-6.83764$ & $26.75616$ & $-47.16495$ & $36.18925$ & $-8.60834$ \\
\hline \hline
\end{tabular}
\vspace{0.5cm}
\caption{\small $E_c$ and $\phi_o$ are in units of eV.  $r_o$ is in 
units of $\AA$, and the other parameters are dimensionless.}
    \label{table:Pparams}
\end{table}
\noindent
\begin{multicols}{2}
\fi
In repeating their melting 
point determination, we found that the value they obtained is not quite 
correct, due primarily to the small system size studied and a problem in 
the choice of thermodynamic integration path for the liquid phase.  Our 
melting point determination was performed using the same basic methodology 
as Mei and Davenport, described below. 

The Gibbs free energies per particle of the liquid and crystal as a 
function of temperature at constant pressure can be obtained by 
thermodynamic integration using  the integral form of the Gibbs-Helmholtz 
equation
\begin{equation} 
\frac{g(T,P)}{T} = \frac{g(T_0,P)}{T_0} - \int_{T_0}^{T}
   \frac{e(\tau,P)}{\tau^2} d \tau      \; ,       \label{gibbshelm}
\end{equation}
where $T_0$ is a predetermined reference temperature and $e(T,P)$ is the 
average total energy per particle.  The Gibbs free energy at the reference 
temperature must be obtained separately by thermodynamic integration from 
a suitable ideal reference state. To do this, the interaction potential 
is parameterized along a linear path between that of the reference potential, 
${\cal U}_0$ and the full potential ${\cal U}$
\begin{equation}
{\cal U}(\xi) = \xi {\cal U} + (1 - \xi) {\cal U}_{0} \;, \label{solidlam}
\end{equation}
The Gibbs free energy per particle relative to that of the reference system
can then be obtained by thermodynamic integration along the path: 
\ifpreprintsty
\begin{equation}
g(T,P) \equiv g(T,P;\xi=1) = g(T_0,P;\xi=0) + \int_0^1 d\xi \left < 
   \frac{\partial e(\xi)} {\partial \xi} \right>_\xi \; \label{solfree}.
\end{equation}
\else
\begin{eqnarray}
g(T,P) & & \; \equiv g(T,P;\xi=1) \nonumber \\
       & & \; = g(T_0,P;\xi=0) + \int_0^1 d\xi \left < 
   \frac{\partial e(\xi)} {\partial \xi} \right>_\xi \; \label{solfree}.
\end{eqnarray}
\fi

For the crystal, the reference system chosen is that of an Einstein 
crystal\cite{Frenkel97}
\begin{equation}
{\cal U}_E(\{r_i\}) = \frac{1}{2} m \omega_{D}^{2} \sum_{i} ({\bf r}_i - 
{\bf r}_{i0})^2 \; ,
\end{equation}
where the $\{{\bf r}_{i0} \}$ are the ideal crystal lattice positions, 
$m$ is the mass and $\omega_D$ is the Debye frequency, which to 
minimize the difference between the reference and full system, should be
chosen to give a similar mean squared displacement for the atoms at the 
temperature of interest.  For this system at 296K, the optimum Debye frequency 
corresponds to a Debye temperature ($T_D=\hbar \omega_D /k$) of 207K. Note 
that this frequency is different than the one used by Mei and Davenport.  

The reference system for the liquid phase is an ideal gas, but the 
transformation must performed as a two step process in order to avoid 
the liquid/gas phase transition.  Mei and Davenport follow the reversible
expansion method used by Broughton and Li\cite{Broughton87b}.  The two 
step process is accomplished by first turning off the attractive part of 
the potential followed a volume expansion to reach the ideal gas limit.
In step one, it is extremely important to turn off the attractive part
of the potential in a way that will not drastically alter the effective
radius of the potential.  If care is not taken, the system will freeze
during the integration of this step.  Here, we write the interatomic 
potential as a linear interpolation between the actual potential and 
the reference system 
\begin{eqnarray}
\phi(r;\zeta) &=& \phi_{rep}(r)  + \zeta \; \phi_{att}(r) \label{liqlam} \;.
\end{eqnarray}
As the value of $\zeta$ is varied from 1 to 0, the system is transformed
from the original Mei and Davenport potential to that of a purely
repulsive potential.  
Step two of the integration is a volume expansion. The free energy
change in this step is given by
\begin{equation}
\Delta g_{\mbox{step 2}} = k_b \, T_0 \int_{0}^{\rho} \frac{d\rho^\prime}
  {\rho^\prime} \left[ \frac{\beta P}{\rho^\prime} - 1 \right] \;. 
  \label{step2int}
\end{equation}
The potential splitting that Mei and Davenport use has the problem in 
that their repulsive potential $\phi_{rep} = \phi_0 \delta \exp{\left[ 
-\gamma \left( \frac{r}{r_0} - 1 \right) \right]}$ has an effective radius 
that is much larger than the actual potential (See Fig.~\ref{fig:Meiphi}a) 
so that as the rest of the potential is turned off the system freezes as 
the effective packing fraction increases.  To remedy this we use a 
WCA\cite{Weeks71} splitting where the repulsive part of the potential 
is equal to the the potential energy for radii less than the radius at 
the minimum of the potential and zero for larger radii. This prescription 
(illustrated in Fig.~\ref{fig:Meiphi}) gives an effective radius more 
similar to the full system and avoids the freezing transition. 

\ifpreprintsty
\else
\begin{figure}
\epsfxsize = 12.5 cm
\epsfbox{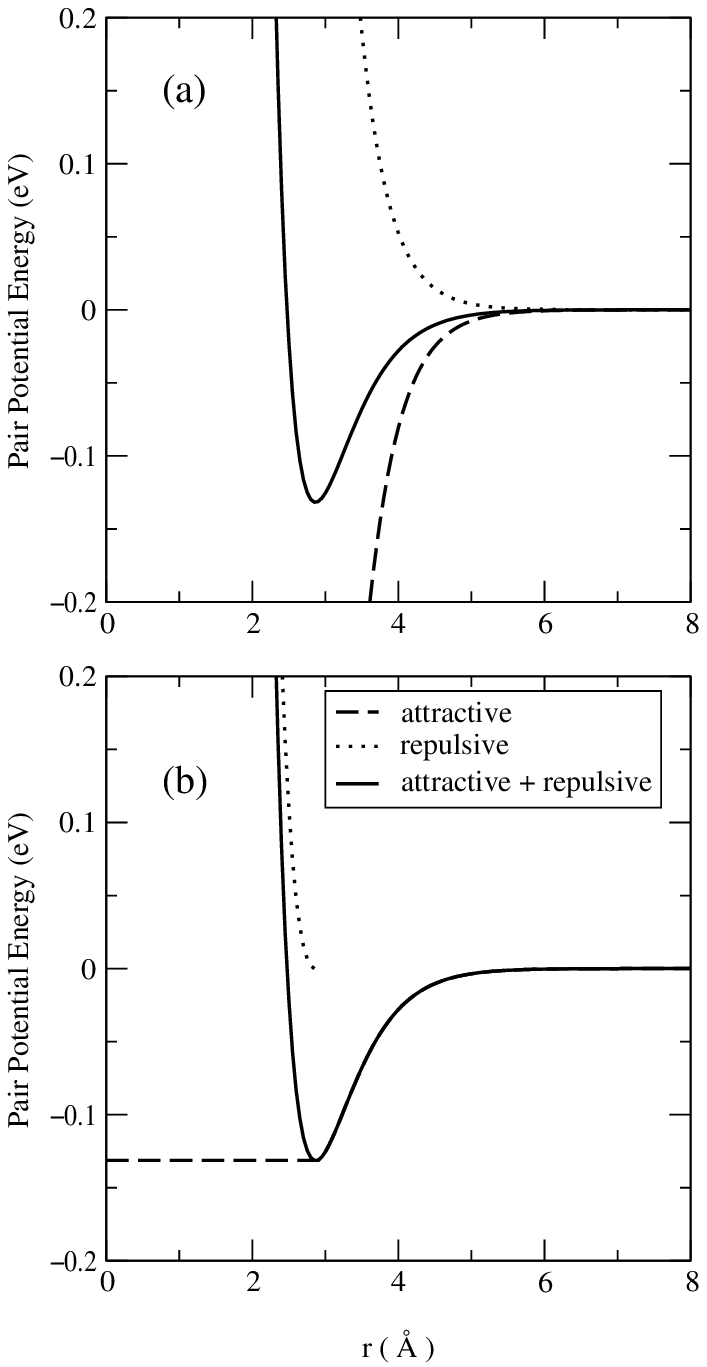}
\vspace{0.5cm}
\caption{\small The different splitting methods for the pair part of 
the aluminum embedded atom potential are graphed separately as a 
function of $r$.  (a) Mei and Davenport splitting (b) WCA splitting
method.  Note the difference in the {\em effective} radii between 
the combined and repulsive parts of the potential.  In the Mei and
Davenport splitting this difference is large and can lead to freezing
during the course of integration, whereas the WCA method has a much 
smaller difference in {\em effective} radii.}
                                                  \label{fig:Meiphi}
\noindent
\end{figure}   
\fi

In order to obtain energy curves needed in Eq.~\ref{gibbshelm} as a 
function of temperature, molecular dynamics (MD) simulations were 
conducted at several different temperatures using the Nos\'e-Poincar\'e-Anderson 
algorithm~\cite{Sturgeon00} for isothermal-isobaric molecular dynamics.  
The fcc crystal was simulated at 50K intervals from 296K to 946K, while 
the liquid was studied over a smaller temperature range from 762K to 1152K 
using 30K intervals.  All of the isothermal-isobaric MD simulation runs 
were equilibrated for 100,000 steps and sampled for 300,000 steps.  From 
these simulations we obtained the average energy and density for both the 
crystal and liquid as a function of temperature and system size.  Both 
the energies and densities were fit to second order polynomials.  The 
coefficients for these polynomials are shown in Table~\ref{table:Ecoeff}.  
These polynomials were used in the construction of the free energy curves 
as described in Eq.~\ref{gibbshelm}.

\ifpreprintsty
\else
\begin{figure}
\epsfxsize = 10 cm
\epsfbox{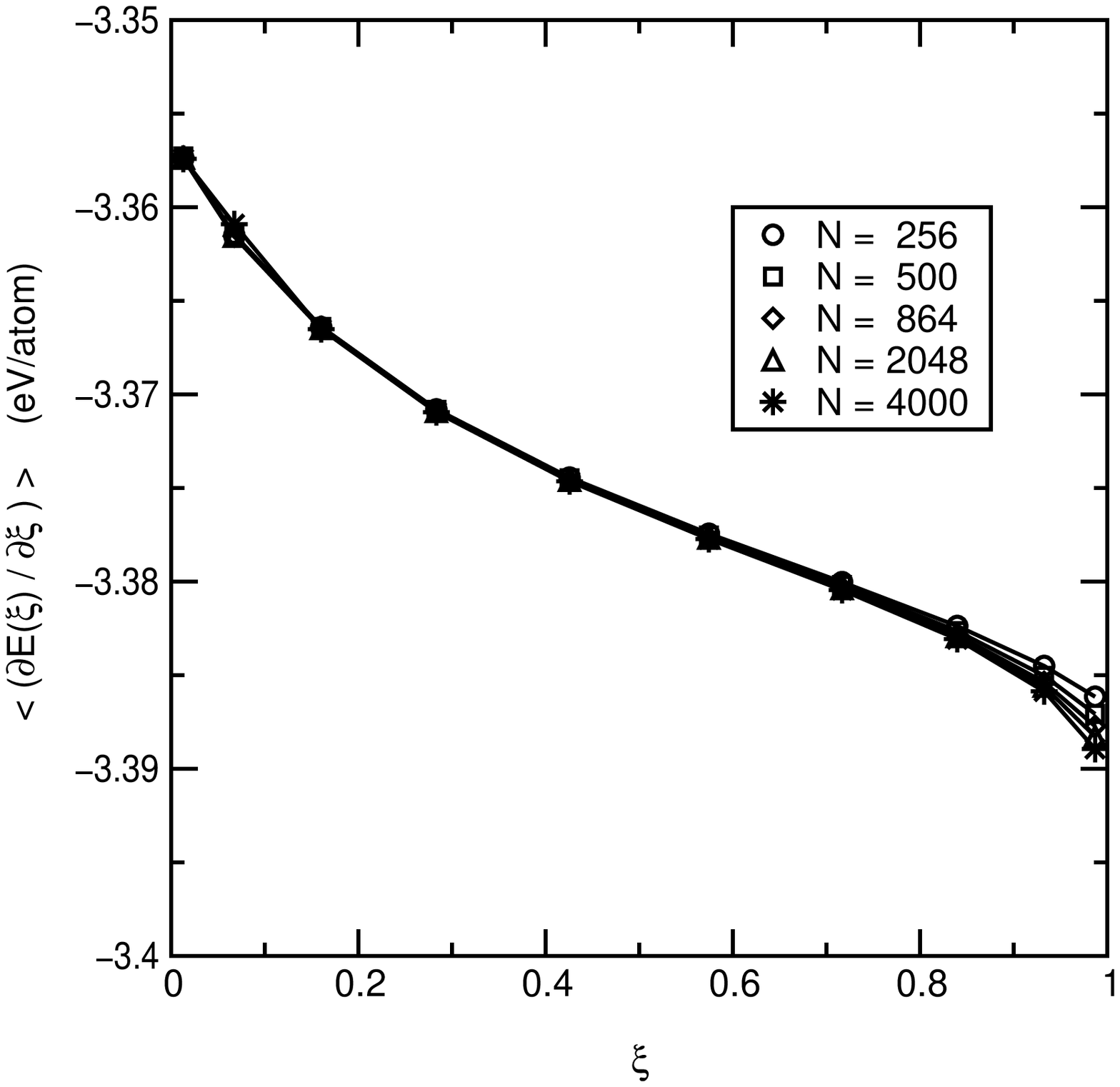}
\vspace{0.5cm}
\caption{\small Integrand for the numerical integration for the solid free 
energy at 296$K$ as a function of $\xi$.  Values of $\xi$ were chose based 
on the ten-point Gauss-Legendre quadrature.  As the value of $\xi$ changes 
from 1 to 0, the system is transformed from an embedded atom solid to that 
of an Einstein crystal.} \label{fig:solf}
\noindent
\end{figure}
\fi
Using the density obtained from the constant $NPT$
simulations, the Gibbs free energy of the fcc crystal at $T_0 = 296K$ was 
calculated by running several simulations at constant N,V,T using the Einstein 
crystal reference state to evaluate the integrand of Eq.~\ref{solfree} at 
values of $\xi$ corresponding to a ten-point Gauss-Legendre quadrature. The 
simulations were performed using the Nos\'e-Poincar\'e algorithm~\cite{Bond99}. 
Figure~\ref{fig:solf} shows a plot of the integrand\  vs.  $\xi$ for all of 
the system sizes studied.  At a value of $\xi = 1 \,$ the system is governed 
solely by the 
\ifpreprintsty
\else
\end{multicols}
\begin{table}
\bf Table~\ref{table:Ecoeff}:  Coefficients for the Energy and Density Polynomials 
\vspace{0.5cm}
\begin{tabular}{r@{\hspace{0.45cm}}|@{\hspace{0.45cm}}c@{\hspace{0.45cm}}
   c@{\hspace{0.45cm}}c@{\hspace{0.45cm}}c@{\hspace{0.45cm}}c@{\hspace{0.45cm}}}
N $\;$ & \multicolumn{1}{c}{256} & \multicolumn{1}{c}{500} & \multicolumn{1}{c}{864} 
  & \multicolumn{1}{c}{2048} &  \multicolumn{1}{c}{4000} \\ 
\hline \hline
Liquid Energy  
$a_0$ & $-3.95773\times 10^{-8}$ & $-2.62765\times 10^{-8} $ & $-3.47219\times 
  10^{-8}$ & $-2.82724\times 10^{-8}$ & $-3.14663\times 10^{-8}$\\
$a_1$ & $ 4.24388\times 10^{-4}$ & $ 3.98266\times 10^{-4} $ & $ 4.15658\times 
  10^{-4}$ & $ 4.02841\times 10^{-4}$ & $ 4.08950\times 10^{-4}$\\
$a_2$ & $-3.39543$ & $-3.38194 $ & $-3.39085$ & $-3.38448$ & $-3.38731$\\
Solid Energy 
$a_0$ & $ 5.75263\times 10^{-8}$ & $ 5.85627\times 10^{-8} $ & $ 5.88951\times 
  10^{-8}$ & $ 5.88624\times 10^{-8}$ & $ 5.88849\times 10^{-8}$\\
$a_1$ & $ 2.30701\times 10^{-4}$ & $ 2.30331\times 10^{-4} $ & $ 2.30283\times 
  10^{-4}$ & $ 2.30528\times 10^{-4}$ & $ 2.30565\times 10^{-4}$\\
$a_2$ & $-3.38550$ & $-3.38533 $ & $-3.38529 $ & $-3.38532 $ & $-3.38531 $\\
\hline
Liquid Density 
$a_0$ & $ 7.59175\times 10^{-10}$ & $ 1.61803\times 10^{-10}$ & $ 5.35909\times 
  10^{-10}$ & $ 3.19601\times 10^{-10}$ & $ 3.91853\times 10^{-10}$\\ 
$a_1$ & $-1.03177\times 10^{-5}$ & $-9.15213\times 10^{-6}$ & $-9.91989\times 
  10^{-6}$ & $-9.47974\times 10^{-6}$ & $-9.62320\times 10^{-6}$\\
$a_2$ & $ 6.09309\times 10^{-2}$ & $ 6.03495\times 10^{-2}$ & $ 6.07444\times 
  10^{-2}$ & $ 6.05232\times 10^{-2}$ & $ 6.05920\times 10^{-2}$\\
Solid Density 
$a_0$ & $-2.59267\times 10^{-9}$ & $-2.62088\times 10^{-9}$ & $-2.63749\times 
  10^{-9}$ & $-2.64149\times 10^{-9}$ & $-2.64086\times 10^{-9}$\\ 
$a_1$ & $-2.79768\times 10^{-6}$ & $-2.78739\times 10^{-6}$ & $-2.77752\times 
  10^{-6}$ & $-2.77869\times 10^{-6}$ & $-2.78125\times 10^{-6}$\\
$a_2$ & $ 6.00427\times 10^{-2}$ & $ 6.00404\times 10^{-2}$ & $ 6.00379\times 
  10^{-2}$ & $ 6.00380\times 10^{-2}$ & $ 6.00387\times 10^{-2}$\\
\end{tabular}
\vspace{0.5cm}
\caption{\small Coefficients for the solid and liquid energy and 
density curves.  The curves are polynomials of the form: $y = a_0 
x^2 + a_1 x + a_2$.} \label{table:Ecoeff}
\end{table}
\begin{multicols}{2}
\noindent
\fi
embedded atom potential. As the value of $\xi$ goes to zero, 
the potential is gradually changed to that of an Einstein crystal.  For each 
value of $\xi$, the system was equilibrated for 100,000 steps and sampled for 
100,000 steps.

Simulations for step one of the liquid free-energy at a temperature of
$1092K$ were performed in the same manner as for the crystal.  Here 
the attractive part of the potential (Eq.~\ref{liqlam}) was slowly turned 
off as the value $\xi$ was changed from one to zero.  
This is shown graphically in Figure~\ref{fig:liqfs}(a).  As with the solid, 
ten-point Gauss-Legendre integration was used to numerically compute this 
integral.  The second step of the liquid free-energy calculation included 
a series of constant $NVT$ simulations at decreasing densities starting with 
the repulsive potential system from the conclusion of step one.  Average 
values for the step two integrand (Eq.~\ref{step2int}) are shown 
graphically in Figure~\ref{fig:liqfs}(b).  The simulations for both 
steps of the liquid free energy integrations were equilibrated for 
100,000 steps and sampled for 100,000 steps at each value of the 
integrand.  The numerical integration for step two was performed using 
the ten-point Simpsons quadrature.  This method was chosen over the 
Gauss-Legendre quadrature due to the inaccuracy of the sampling at low 
density.  The value of the integrand at zero density was obtained by 
an analytical calculation of the second virial coefficient ($B_2$).  

Free energies and the melting point are shown in Table~\ref{table:meltT} as a 
function of particle number.  A graph of the 
\ifpreprintsty
\else
\end{multicols}
\begin{table} 
\bf Table~\ref{table:meltT}:  Free Energies and Melting Points for 
    $\delta = 7.35$
\vspace{0.5cm}
\begin{tabular}{r@{\hspace{0.45cm}}|@{\hspace{0.45cm}}r@{\hspace{0.45cm}}
  r@{\hspace{0.45cm}}r@{\hspace{0.45cm}}r@{\hspace{0.45cm}}r@{\hspace{0.45cm}}}
N $\;$ & \multicolumn{1}{c}{256} & \multicolumn{1}{c}{500} 
  & \multicolumn{1}{c}{864} & \multicolumn{1}{c}{2048}& \multicolumn{1}{c}{4000} \\ 
\hline \hline
Original $\delta \;$ & \multicolumn{1}{c}{7.35} & \multicolumn{1}{c}{7.35} 
  & \multicolumn{1}{c}{7.35} & \multicolumn{1}{c}{7.35}& \multicolumn{1}{c}{7.35} \\ 
\hline
$g_{s}(T = 296K) \;$ (eV/atom) & -3.40200(3) & -3.40219(2) & -3.40236(2) & -3.40246(1) 
  & -3.40254(1)\\
$g_{l}(T = 1092K) \;$ (eV/atom) & -3.8561(6)  & -3.8561(4)  & -3.8563(3)  & -3.8565(2)  
  & -3.8565(2) \\
$T_{m} \;$ (K) & 802.8$\pm$5.6  & 814.3$\pm$ 3.7  & 819.7$\pm$2.9 & 822.5$\pm$1.9 
  & 825.2$\pm$1.3 \\
\hline \hline
New $\delta \;$ & \multicolumn{1}{c}{8.70} & \multicolumn{1}{c}{8.57} 
  & \multicolumn{1}{c}{8.50} & \multicolumn{1}{c}{8.47}& \multicolumn{1}{c}{8.45} \\ 
\hline
$g_{s}(T = 296K) \;$ (eV/atom) & $-3.39538(3)$ & $-3.39615(2)$ & $-3.39664(1)$ 
  & $-3.39687(1)$ & $-3.39704(1)$\\
$g_{l}(T = 1092K) \;$ (eV/atom) & $-3.8135(6)$  & $-3.8175(5)$  & $-3.8200(3) $ 
  & $-3.8209(2)$  & $-3.8217(2)$ \\
$T_{m} \;$ (K) & $934.9 \pm 5.9$  & $933.1 \pm 4.4$  & $930.9 \pm 3.0$ 
  & $931.9 \pm 2.0$ & $931.5 \pm 1.5$ \\
\end{tabular}
\vspace{0.5cm}
\caption{\small Simulation averages for the solid and liquid Gibbs free 
energies along with the calculated melting temperature for several  
system sizes.   Free energies are in units of $eV/atom$ 
and the melting temperature is in $K$.} \label{table:meltT}
\end{table}
\noindent
\begin{multicols}{2}
\fi
melting temperature vs. $1/N$ 
(Figure~\ref{fig:meltT}) shows that at infinite particle number the melting 
point for the embedded atom potential proposed by Mei and Davenport approaches 
$826.4K \pm 1.3K$. It should be noted that the major error in the melting point 
calculated by Mei and Davenport for their potential 800$\pm 9K$ is primarily 
due to the small size of their system. The problems with their potential 
splitting appears to have had little effect, probably due to cancellation of 
errors, as the melting point that we determine here for the 256 particle system 
agrees with theirs within the simulation error. Recently, Morris, Wang, Ho 
and Chan~\cite{Morris94} argued on the basis of the stability of crystal-liquid 
interfaces that the melting point of the Mei-Davenport aluminum potential is 
actually significantly lower (around 725K). Our results do not support this 
conclusion and as a check we have carefully set up stable stress-free interfaces 
at our calculated melting point. The systems exhibit melting (freezing) as the 
temperature is raised (lowered) away from our calculated melting point.  The 
lower temperature transitions of Morris, {\it et al.} were most likely due to 
significant un-relaxed stress in the crystal. 

\section{Results for a model of Aluminum: Re-parameterizing the potential} 
  \label{sec:Results2}

To adjust the melting temperature of the embedded atom potential for 
aluminum, we first calculate $\frac{\partial T_m}{\partial X_i}$ for each 
of the parameters in the potential. (Note, for simplicity and consistency 
the expansion coefficients, $c_l$, in Eq.~\ref{fexp} were kept constant).  
The derivatives were calculated in a single (constant $NPT$) simulation at 
the melting temperature calculated in the previous section with $P=0$, for 
each of several system sizes ($N=$256, 500, 864, 2048 and 4000).  The system 
was equilibrated for 100,000 steps followed by 100,000 steps for averageing.  
From this simulation it was determined that $\frac{\partial T_m}{\partial X_i}$ 
was significant only for the parameter $\delta$ - the other parameters of the 
potential have little effect on the melting point. The complicated nature of 
the potential makes it difficult to assign any physical explanation to the 
sensitivity of the melting point to $\delta$ relative to the other parameters.  

Next, a series of simulations 
were performed for each system size to integrate along the coexistence 
curve from the initial calculated melting point of the potential to the 
true experimental melting temperature.  At each temperature along the 
coexistence curve, the system was equilibrated for 100,000 steps and 
sampled over 300,000 steps.  In our experiments we use a fourth-order 
predictor corrector integrator to carry out the integration along the 
coexistence curve as a function of $\delta$.  Fig~\ref{fig:dint} shows 
this integration graphically with final values of $\delta$ corresponding 
to a melting temperature of $T_m = 933K$.  These values are listed in 
Table~\ref{table:meltT}. (Note that, although we did an accurate
numerical integration along the coexistence 
\ifpreprintsty
\else
\begin{figure}
\epsfxsize = 12.5 cm
\epsfbox{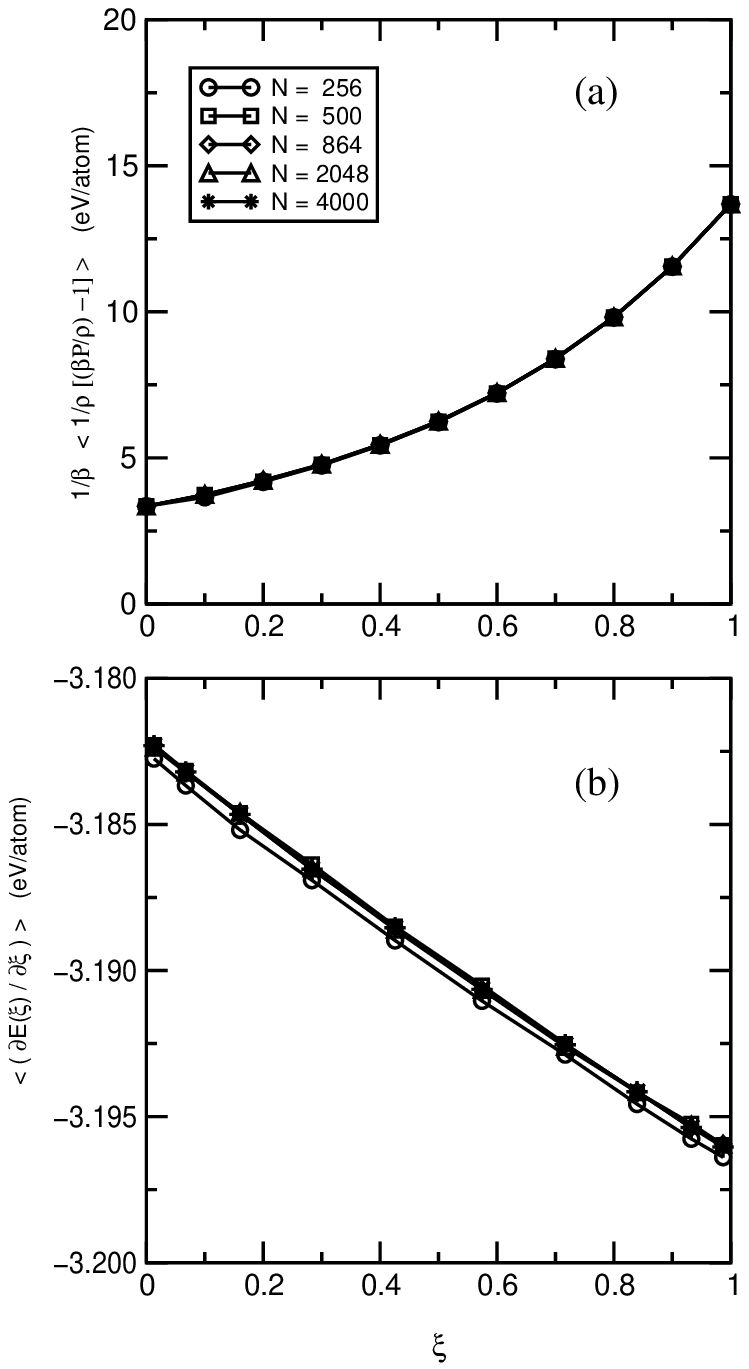}
\caption{\small {\it (a)} Simulation results for the integrand of step 
one of the liquid free energy calculation  at $T = 1092K$ as a function of $\xi$.  This 
integration slowly turns off the attractive part of the potential as the
 value of $\xi$ changes from 1 to 0.  Values for $\xi$ were again based 
upon the ten-point Gauss-Legendre quadrature. {\it (b)}
Integrand for the volume expansion integration (step 2) in  
the liquid free energy calculation at $T = 1092K$ as a function of 
$\rho^*  = \frac{\rho}{\rho_0}$.  The numerical integration for this
step was performed using the ten-point Simpson quadrature.  The value 
at $\rho^* = 0$ was obtained by an analytic calculation of the second 
virial coefficient ($B_2(T)$).} \label{fig:liqfs}
\end{figure}   
\noindent
\fi
curve, the results indicate that using the approximation that the deriviative 
is a constant in the region of interest would have been correct to within 
the simulation error.)

\ifpreprintsty
\else
\begin{figure}
\epsfxsize = 10 cm
\epsfbox{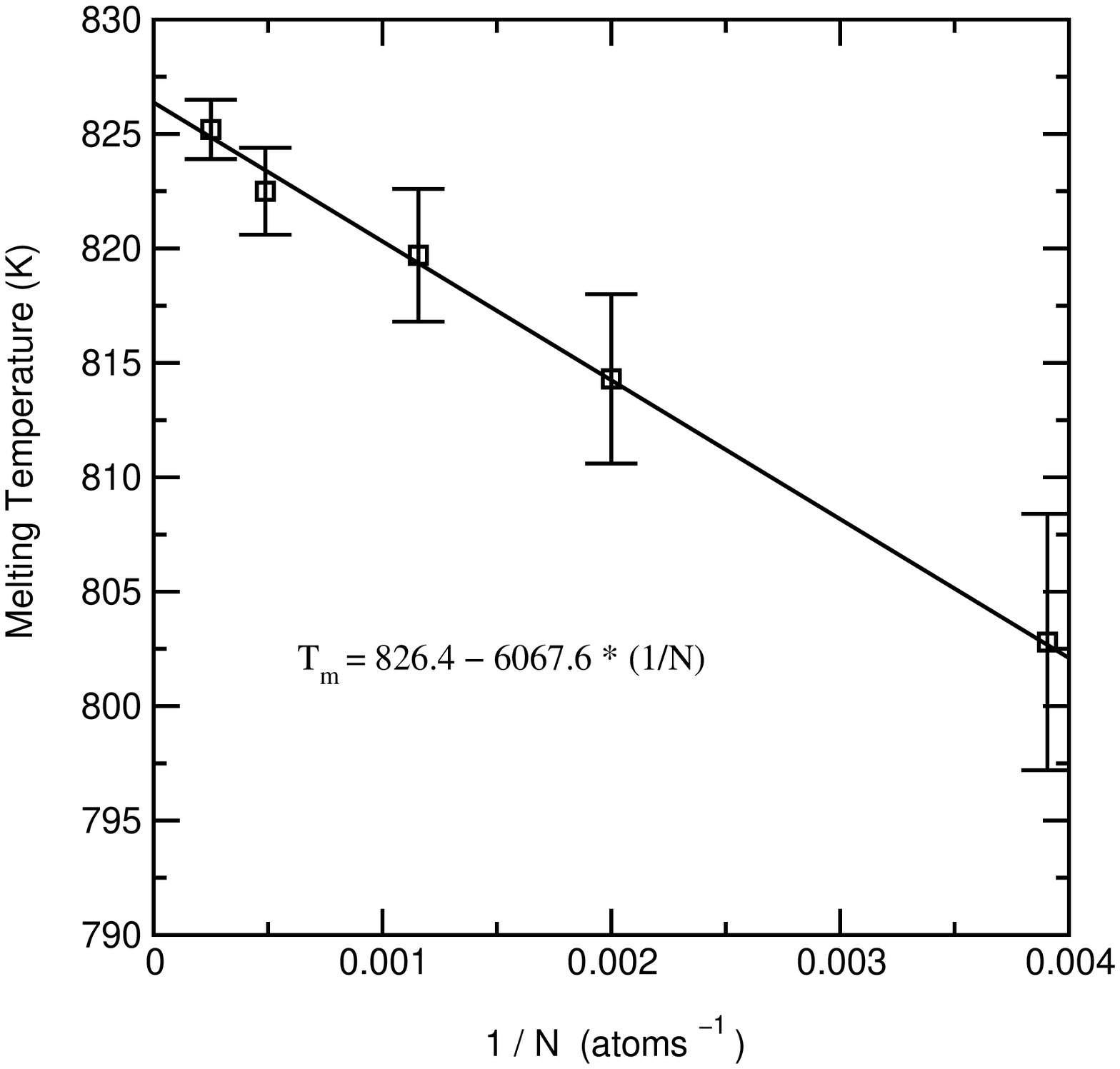}
\caption{\small The melting temperature of the embedded atom potential 
is plotted as a function of inverse particle size.  As $N$ goes to
infinity ($\frac{1}{N} \rightarrow 0$), the melting temperature 
approaches $826.4K$.}
                                                  \label{fig:meltT}
\end{figure}   
\noindent
\fi

In order to confirm that the melting point did indeed change as expected, 
the melting point calculation was repeated using the newly calculated value 
of $\delta$.  The new melting temperature for the embedded atom potential 
with a value of $\delta=8.45$ corresponding to a system size of N=4000, 
was calculated to be $931.5K \pm 1.5K$.  Melting temperatures for the
other system sizes are listed in Table~\ref{table:meltT}.  The experimental 
value for the melting temperature of aluminum is 933.47K.

Mei and Davenport initially determined $\delta$ (and the other parameters 
in the potential) by fitting the potential to certain physical constants.  
For the new value of $\delta$ (8.45), we have re-calculated a variety of 
physical properties of aluminum.  These quantities, for the original potential, 
the reparameterized potential and the experimental values are collected in 
Table~\ref{table:c11}.  From this data, 
\ifpreprintsty
\else
\begin{figure}
\epsfxsize = 10 cm
\epsfbox{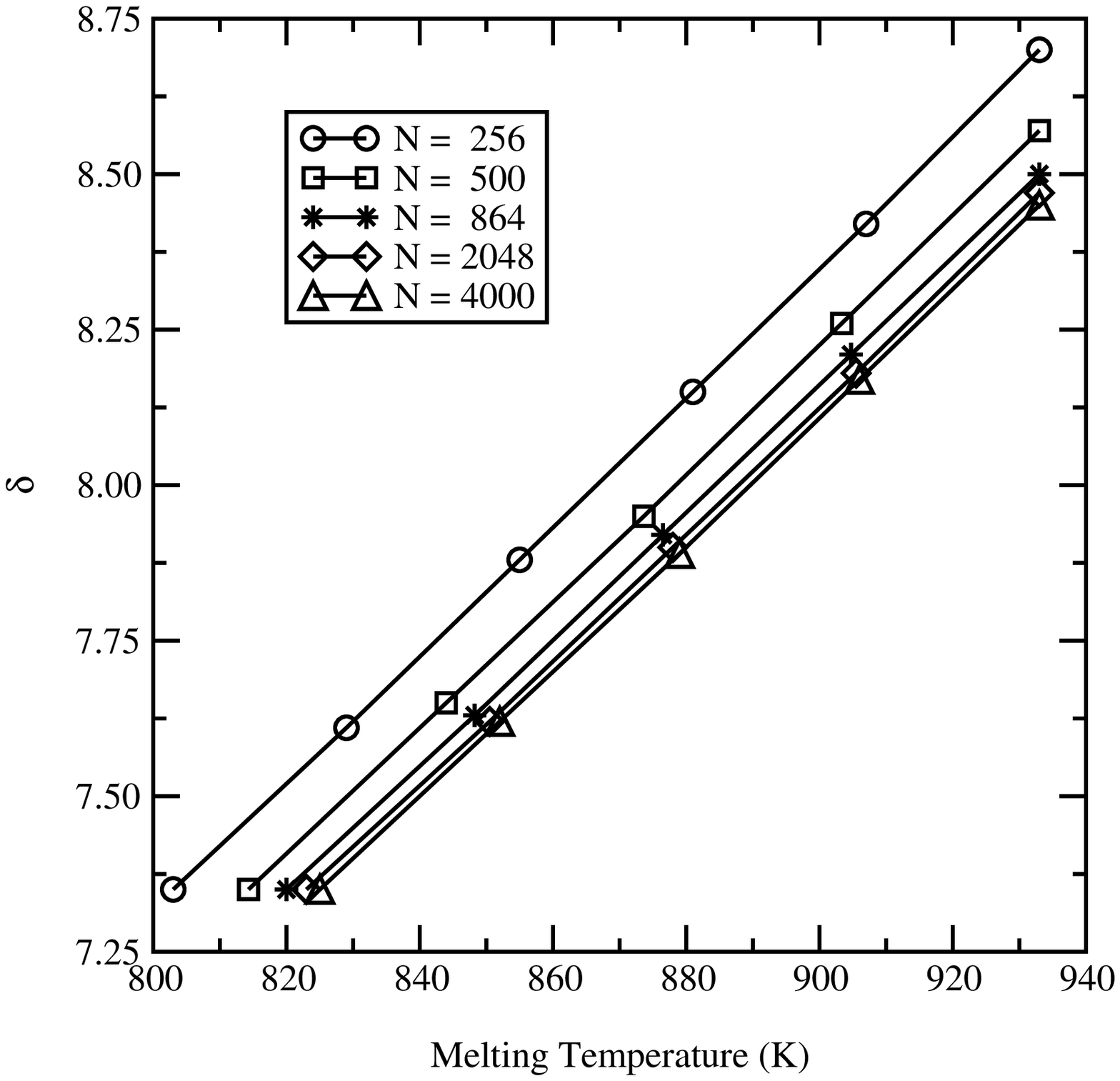}
\caption{\small Values of $\delta$ along the Gibbs-Duhem integration 
of the crystal-melt coexistence curve.  These curves were generated 
using a fourth-order predictor-corrector and show the melting 
temperature as a function of $\delta$ for each of the various system 
sizes.} \label{fig:dint}
\end{figure}   
\noindent
\fi
\ifpreprintsty
\else
\end{multicols}
\begin{table} 
\bf Table~\ref{table:c11}:  Physical constants for N=4000
\vspace{0.5cm}
\begin{tabular}{r@{\hspace{0.45cm}}|@{\hspace{0.45cm}}r@{\hspace{0.45cm}}
  r@{\hspace{0.45cm}}r@{\hspace{0.45cm}}r@{\hspace{0.45cm}}}
$\;$ & \multicolumn{1}{c}{$\delta = 7.35$} & \multicolumn{1}{c}{$\delta = 8.45$} 
  & \multicolumn{1}{c}{Experimental} \\
\hline \hline
Un-relaxed Vacancy Formation (eV/atom) & 4.07211 & 4.18679 & N/A \\
Latent Heat (eV/atom) & 0.0830 & 0.0973 & 0.111  \\
$C_{11}$ (dynes/cm$^2$)  &   0.93   &   0.96   &   1.07   \\
$C_{12}$ (dynes/cm$^2$)  &   0.69   &   0.68   &   0.61   \\
$C_{44}$ (dynes/cm$^2$)  &   0.33   &   0.36   &   0.28   \\
\end{tabular}
\vspace{0.5cm}
\caption{\small Here the vacancy formation energy, latent heat 
values and elastic constants are presented for the original version 
of the EAM Al potential and the modified version.  $C_{11}$ and $C_{12}$ 
for the modified version (with the experimental melting temperature) 
are closer to their experimental values.  $C_{44}$ seems to get slightly 
worse. } \label{table:c11}
\end{table}
\begin{multicols}{2}
\noindent
\fi
we see that, in comparison to the 
original potential with $\delta = 7.35$, the new potential more closely 
models the experimental values~\cite{Ashcroft88} of the ($T=0K$) elastic 
constants $C_{11}$ and $C_{22}$ while $C_{44}$  becomes slightly worse in 
comparison to its experimentally determined value.  In addition, mostly as 
a consequence of the improved melting point, the latent heat is considerably 
improved. 

\section{Summary} \label{sec:Summary}
We have outlined  an application of the Gibbs-Duhem integration method 
of Kofke\cite{Kofke93a}, with which a model interaction potential can be 
re-parameterized, including the experimental melting point in the optimization 
protocol.  The melting temperature of a potential can then be adjusted similar 
to the tuning of other parameters in the potential.  Since non-mechanical 
properties, such as the melting point, are not generally included
in potential optimization and the agreement of such quantities with experiment
is not guaranteed, such a procedure will be useful in situations, such as in
the simulation of crystal-melt interfaces, where the having the correct melting 
point is highly desirable. The method is general and  can easily be extended to 
a variety of systems. As an example of the utility of the method, we apply the
procedure to reparameterizing a popular model of aluminum\cite{Mei92} for
which the melting point has been calculated to be over 100K below the accepted
experimental value. We demonstrate that the reparamterized potential has
a melting point that agrees within the statistical error with the experimental
value of 933K and that reparameterization does not degrade the quality of 
the potential with respect a variety of properties, and that, in fact, 
quantities such as the elastic constants $C_{11}$ and $C_{12}$ and the 
latent heat, agreement is improved.

\section{Acknowledgements} \label{sec:Acknowledgements}
We gratefully acknowledge J. W. Davenport for helpful conversations,
as well as the Kansas Center for Advanced Scientific Computing for 
the use of their computer facilities. We also would like to thank the 
National Science Foundation (under grant CHE-9500211) as well as the 
University of Kansas General Research Fund for their generous support 
of this research.
\vspace{0.85in} \\

\newpage

\ifpreprintsty
%*********************************************************************
%        FIGURE:  Mei&Davenports split potential
%*********************************************************************
\begin{figure}
\epsfxsize = 8.3 cm
\rput[bl]{-90}(-0.5,3.0){\epsfbox{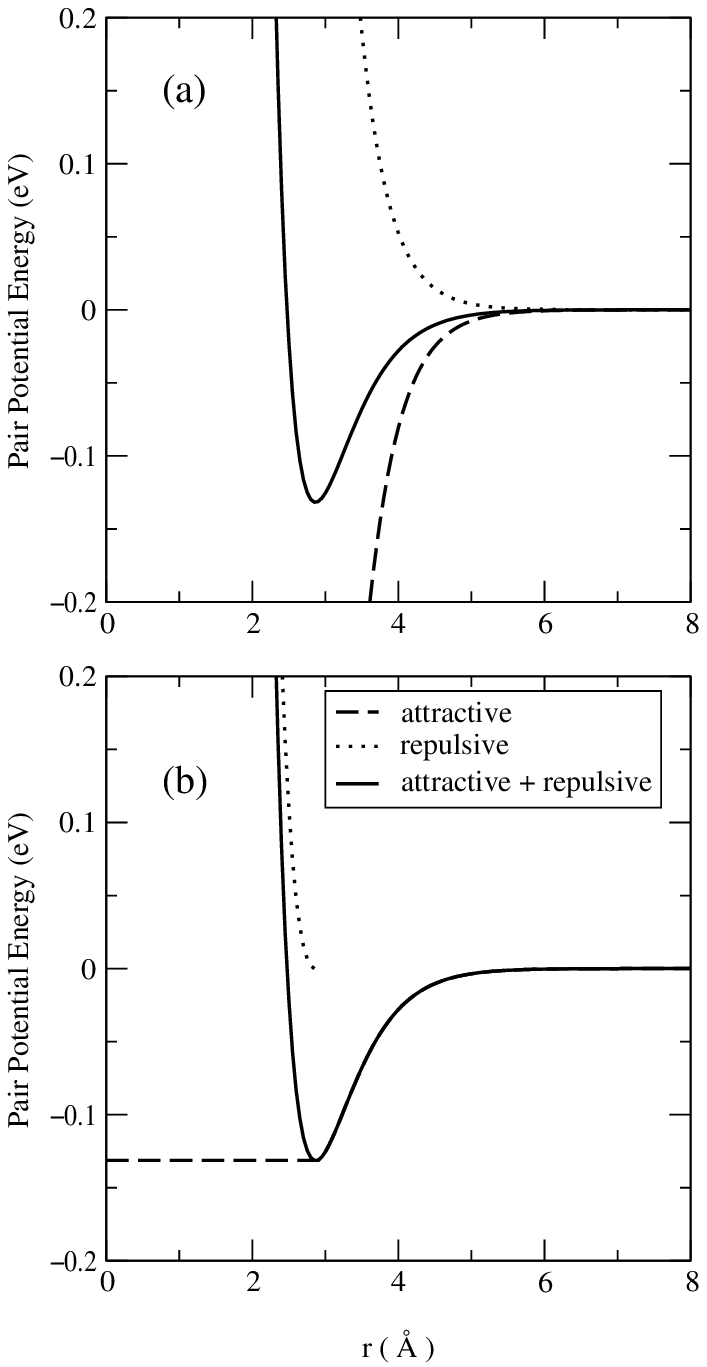}}
\vspace{6.0in}
\caption{\small The different splitting methods for the pair part of 
the aluminum embedded atom potential are graphed separately as a 
function of $r$.  (a) Mei and Davenport splitting (b) WCA splitting
method.  Note the difference in the {\em effective} radii between 
the combined and repulsive parts of the potential.  In the Mei and
Davenport splitting this difference is large and can lead to freezing
during the course of integration, whereas the WCA method has a much 
smaller difference in {\em effective} radii.}
                                                  \label{fig:Meiphi}
\end{figure}   
%*********************************************************************
%        FIGURE:   Solid free energy integration using xi
%*********************************************************************
\begin{figure}
\rput[bl](-0.5,-15.0){\epsfbox{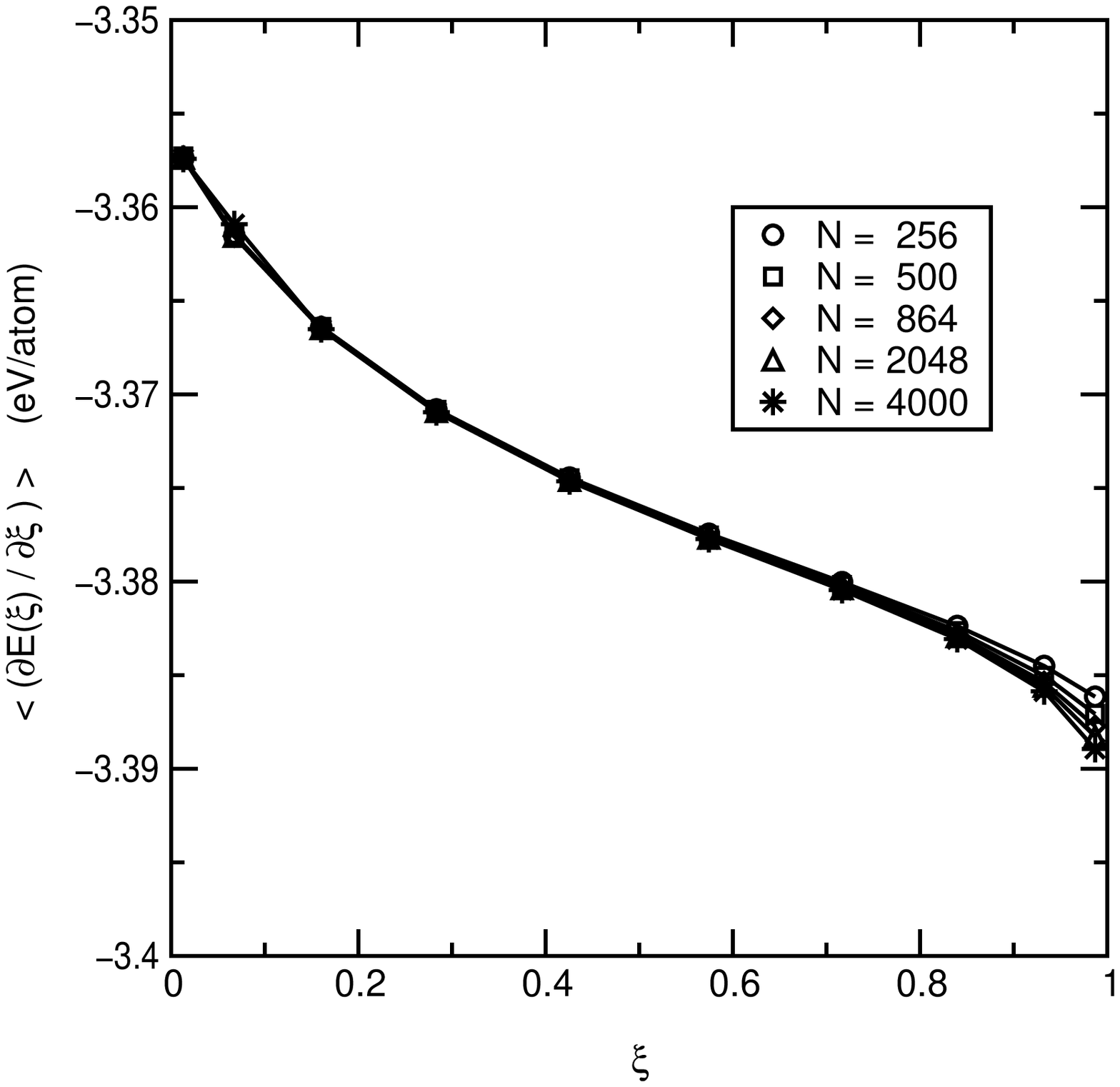}}
\vspace{6.0in}
\caption{\small Integrand for the numerical integration for the solid free 
energy at 296$K$ as a function of $\xi$.  Values of $\xi$ were chose based 
on the ten-point Gauss-Legendre quadrature.  As the value of $\xi$ changes 
from 1 to 0, the system is transformed from an embedded atom solid to that 
of an Einstein crystal.}
                                                  \label{fig:solf}
\end{figure}   
%*********************************************************************
%        FIGURE:   Liquid free-energy integration using xi (step1)
%*********************************************************************
\begin{figure}
\rput[bl]{-90}(-0.5,3.0){\epsfbox{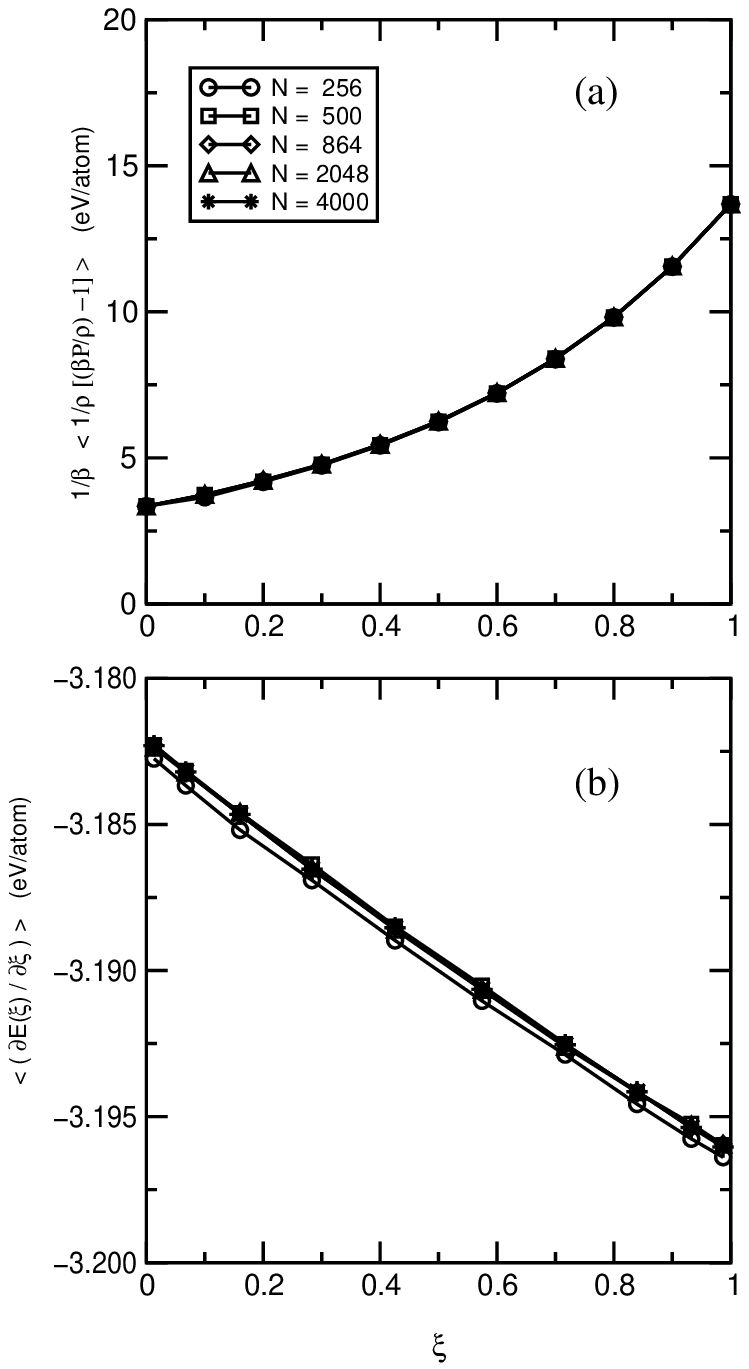}}
\rput[l]( 3.0,-3.75){\large (a) }
\rput[l](17.5,-3.75){\large (b) }
\rput[l](15.0,-11.75){\small * }
\vspace{6.0in}
\caption{\small {\it (a)} Simulation results for the integrand of step 
one of the liquid free energy calculation  at $T = 1092K$ as a function of $\xi$.  This 
integration slowly turns off the attractive part of the potential as the
 value of $\xi$ changes from 1 to 0.  Values for $\xi$ were again based 
upon the ten-point Gauss-Legendre quadrature. {\it (b)}
Integrand for the volume expansion integration (step 2) in  
the liquid free energy calculation at $T = 1092K$ as a function of 
$\rho^*  = \frac{\rho}{\rho_0}$.  The numerical integration for this
step was performed using the ten-point Simpson quadrature.  The value 
at $\rho^* = 0$ was obtained by an analytic calculation of the second 
virial coefficient ($B_2(T)$).}
                                                  \label{fig:liqfs}
\end{figure}   
%*********************************************************************
%        FIGURE:   Melting Temperature vs. 1/N
%*********************************************************************
\begin{figure}
\rput[bl](-0.5,-15.0){\epsfbox{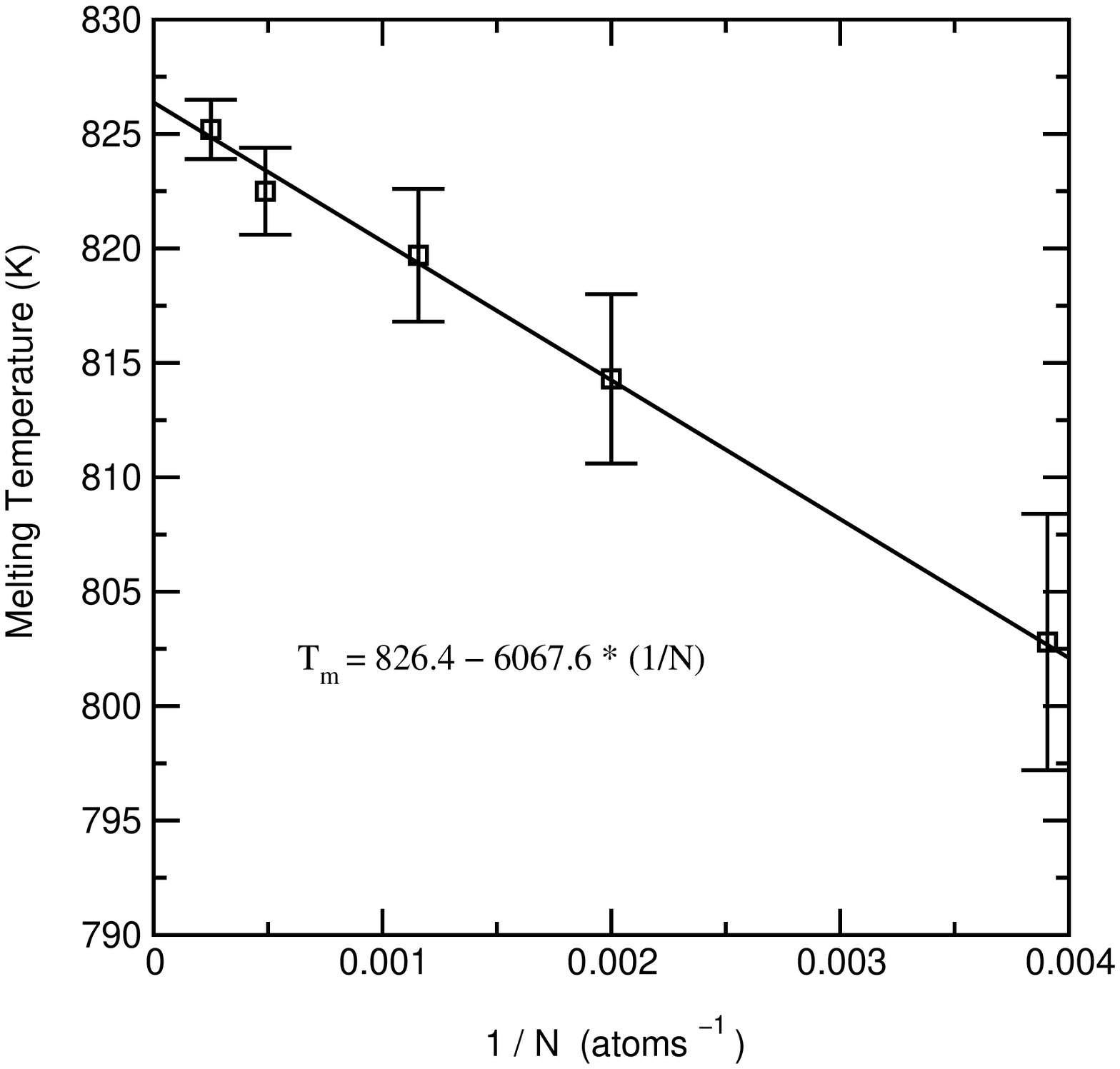}}
\vspace{6.0in}
\caption{\small The melting temperature of the embedded atom potential 
is plotted as a function of inverse particle size.  As $N$ goes to
infinity ($\frac{1}{N} \rightarrow 0$), the melting temperature 
approaches $826.4K$.}
                                                  \label{fig:meltT}
\end{figure}   
%*********************************************************************
%        FIGURE:   Delta Integration
%*********************************************************************
\begin{figure}
\rput[bl](-0.5,-15.0){\epsfbox{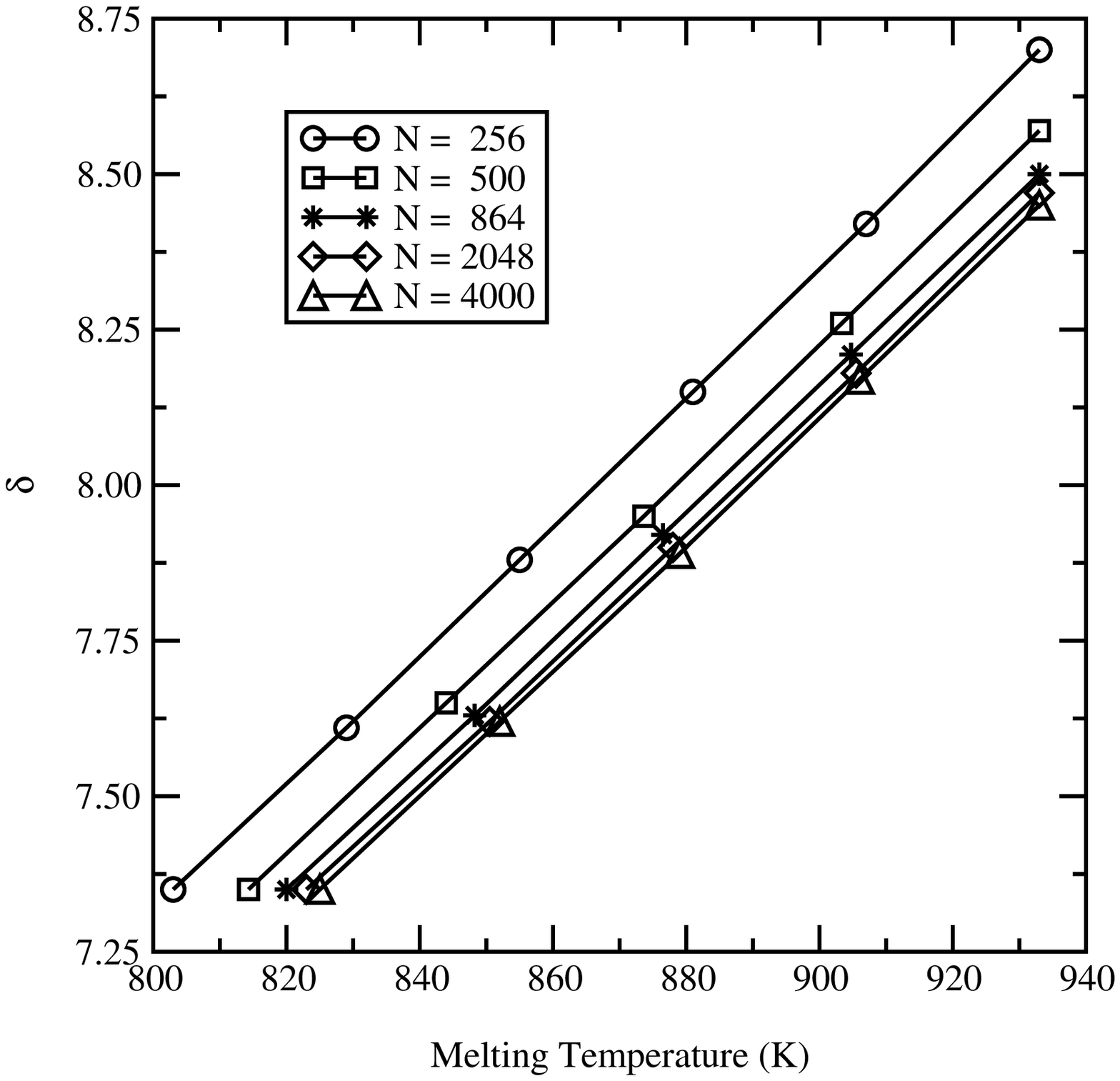}}
\vspace{6.0in}
\caption{\small Values of $\delta$ along the Gibbs-Duhem integration 
of the crystal-melt coexistence curve.  These curves were generated 
using a fourth-order predictor-corrector and show the melting 
temperature as a function of $\delta$ for each of the various system 
sizes.}
                                                  \label{fig:dint}
\end{figure}   
%*********************************************************************
%         TABLE:   Potential parameters for the EAM aluminum of M&D
%*********************************************************************
\begin{table}
\bf Table~\ref{table:Pparams}:  Parameters for the Aluminum embedded
atom potential developed by Mei and Davenport.  
\vspace{0.5cm}
\begin{tabular}{c@{\hspace{0.35cm}}c@{\hspace{0.35cm}}c@{\hspace{0.35cm}}
  c@{\hspace{0.35cm}}c@{\hspace{0.35cm}}c@{\hspace{0.35cm}}c@{\hspace{0.35cm}}}
\hline \hline
$E_c$ & $\phi_0$ & $r_0$ & $\alpha$ & $\beta$ & $\gamma$ & $\delta$ \\
\hline
$3.39$ & $0.1318$ & $2.8638$ & $4.60$ & $7.10$ & $7.34759$ & $7.35$ \\
$c_0$ & $c_1$ & $c_2$ & $c_3$ & $c_4$ & $c_5$ \\
$0.64085$ & $-6.83764$ & $26.75616$ & $-47.16495$ & $36.18925$ & $-8.60834$ \\
\hline \hline
\end{tabular}
\vspace{0.5cm}
\caption{\small $E_c$ and $\phi_o$ are in units of eV.  $r_o$ is in 
units of $\AA$, and the other parameters are dimensionless.}
    \label{table:Pparams}
\end{table}

%*********************************************************************
%         TABLE:   Coefficients for the Energy and Density Polynomials 
%*********************************************************************
\begin{table}
\bf Table~\ref{table:Ecoeff}:  Coefficients for the Energy and Density Polynomials 
\vspace{0.5cm}
\begin{tabular}{r@{\hspace{0.45cm}}|@{\hspace{0.45cm}}c@{\hspace{0.45cm}}
   c@{\hspace{0.45cm}}c@{\hspace{0.45cm}}c@{\hspace{0.45cm}}c@{\hspace{0.45cm}}}
N $\;$ & \multicolumn{1}{c}{256} & \multicolumn{1}{c}{500} & \multicolumn{1}{c}{864} 
  & \multicolumn{1}{c}{2048} &  \multicolumn{1}{c}{4000} \\ 
\hline \hline
Liquid Energy  
$a_0$ & $-3.95773\times 10^{-8}$ & $-2.62765\times 10^{-8} $ & $-3.47219\times 
  10^{-8}$ & $-2.82724\times 10^{-8}$ & $-3.14663\times 10^{-8}$\\
$a_1$ & $ 4.24388\times 10^{-4}$ & $ 3.98266\times 10^{-4} $ & $ 4.15658\times 
  10^{-4}$ & $ 4.02841\times 10^{-4}$ & $ 4.08950\times 10^{-4}$\\
$a_2$ & $-3.39543$ & $-3.38194 $ & $-3.39085$ & $-3.38448$ & $-3.38731$\\
Solid Energy 
$a_0$ & $ 5.75263\times 10^{-8}$ & $ 5.85627\times 10^{-8} $ & $ 5.88951\times 
  10^{-8}$ & $ 5.88624\times 10^{-8}$ & $ 5.88849\times 10^{-8}$\\
$a_1$ & $ 2.30701\times 10^{-4}$ & $ 2.30331\times 10^{-4} $ & $ 2.30283\times 
  10^{-4}$ & $ 2.30528\times 10^{-4}$ & $ 2.30565\times 10^{-4}$\\
$a_2$ & $-3.38550$ & $-3.38533 $ & $-3.38529 $ & $-3.38532 $ & $-3.38531 $\\
\hline
Liquid Density 
$a_0$ & $ 7.59175\times 10^{-10}$ & $ 1.61803\times 10^{-10}$ & $ 5.35909\times 
  10^{-10}$ & $ 3.19601\times 10^{-10}$ & $ 3.91853\times 10^{-10}$\\ 
$a_1$ & $-1.03177\times 10^{-5}$ & $-9.15213\times 10^{-6}$ & $-9.91989\times 
  10^{-6}$ & $-9.47974\times 10^{-6}$ & $-9.62320\times 10^{-6}$\\
$a_2$ & $ 6.09309\times 10^{-2}$ & $ 6.03495\times 10^{-2}$ & $ 6.07444\times 
  10^{-2}$ & $ 6.05232\times 10^{-2}$ & $ 6.05920\times 10^{-2}$\\
Solid Density 
$a_0$ & $-2.59267\times 10^{-9}$ & $-2.62088\times 10^{-9}$ & $-2.63749\times 
  10^{-9}$ & $-2.64149\times 10^{-9}$ & $-2.64086\times 10^{-9}$\\ 
$a_1$ & $-2.79768\times 10^{-6}$ & $-2.78739\times 10^{-6}$ & $-2.77752\times 
  10^{-6}$ & $-2.77869\times 10^{-6}$ & $-2.78125\times 10^{-6}$\\
$a_2$ & $ 6.00427\times 10^{-2}$ & $ 6.00404\times 10^{-2}$ & $ 6.00379\times 
  10^{-2}$ & $ 6.00380\times 10^{-2}$ & $ 6.00387\times 10^{-2}$\\
\end{tabular}
\vspace{0.5cm}
\caption{\small Coefficients for the solid and liquid energy and 
density curves.  The curves are polynomials of the form: $y = a_0 
x^2 + a_1 x + a_2$.} \label{table:Ecoeff}
\end{table}
%*********************************************************************
%         TABLE:   Free Energies and Melting Points D=7.35
%*********************************************************************
\begin{table} 
\bf Table~\ref{table:meltT}:  Free Energies and Melting Points for 
    $\delta = 7.35$
\vspace{0.5cm}
\begin{tabular}{r@{\hspace{0.45cm}}|@{\hspace{0.45cm}}r@{\hspace{0.45cm}}
  r@{\hspace{0.45cm}}r@{\hspace{0.45cm}}r@{\hspace{0.45cm}}r@{\hspace{0.45cm}}}
N $\;$ & \multicolumn{1}{c}{256} & \multicolumn{1}{c}{500} 
  & \multicolumn{1}{c}{864} & \multicolumn{1}{c}{2048}& \multicolumn{1}{c}{4000} \\ 
\hline \hline
Original $\delta \;$ & \multicolumn{1}{c}{7.35} & \multicolumn{1}{c}{7.35} 
  & \multicolumn{1}{c}{7.35} & \multicolumn{1}{c}{7.35}& \multicolumn{1}{c}{7.35} \\ 
\hline
$g_{s}(T = 296K) \;$ (eV/atom) & -3.40200(3) & -3.40219(2) & -3.40236(2) & -3.40246(1) 
  & -3.40254(1)\\
$g_{l}(T = 1092K) \;$ (eV/atom) & -3.8561(6)  & -3.8561(4)  & -3.8563(3)  & -3.8565(2)  
  & -3.8565(2) \\
$T_{m} \;$ (K) & 802.8$\pm$5.6  & 814.3$\pm$ 3.7  & 819.7$\pm$2.9 & 822.5$\pm$1.9 
  & 825.2$\pm$1.3 \\
\hline \hline
New $\delta \;$ & \multicolumn{1}{c}{8.70} & \multicolumn{1}{c}{8.57} 
  & \multicolumn{1}{c}{8.50} & \multicolumn{1}{c}{8.47}& \multicolumn{1}{c}{8.45} \\ 
\hline
$g_{s}(T = 296K) \;$ (eV/atom) & $-3.39538(3)$ & $-3.39615(2)$ & $-3.39664(1)$ 
  & $-3.39687(1)$ & $-3.39704(1)$\\
$g_{l}(T = 1092K) \;$ (eV/atom) & $-3.8135(6)$  & $-3.8175(5)$  & $-3.8200(3) $ 
  & $-3.8209(2)$  & $-3.8217(2)$ \\
$T_{m} \;$ (K) & $934.9 \pm 5.9$  & $933.1 \pm 4.4$  & $930.9 \pm 3.0$ 
  & $931.9 \pm 2.0$ & $931.5 \pm 1.5$ \\
\end{tabular}
\vspace{0.5cm}
\caption{\small Simulation averages for the solid and liquid Gibbs free 
energies along with the calculated melting temperature for several  
system sizes.   Free energies are in units of $eV/atom$ 
and the melting temperature is in $K$.}  
\label{table:meltT}
\end{table}
%*********************************************************************
%         TABLE:   Physical constants and others
%*********************************************************************
\begin{table} 
\bf Table~\ref{table:c11}:  Physical constants for N=4000
\vspace{0.5cm}
\begin{tabular}{r@{\hspace{0.45cm}}|@{\hspace{0.45cm}}r@{\hspace{0.45cm}}
  r@{\hspace{0.45cm}}r@{\hspace{0.45cm}}r@{\hspace{0.45cm}}}
$\;$ & \multicolumn{1}{c}{$\delta = 7.35$} & \multicolumn{1}{c}{$\delta = 8.45$} 
  & \multicolumn{1}{c}{Experimental} \\
\hline \hline
Un-relaxed Vacancy Formation (eV/atom) & 4.07211 & 4.18679 & N/A \\
Latent Heat (eV/atom) & 0.0830 & 0.0973 & 0.111  \\
$C_{11}$ (dynes/cm$^2$)  &   0.93   &   0.96   &   1.07   \\
$C_{12}$ (dynes/cm$^2$)  &   0.69   &   0.68   &   0.61   \\
$C_{44}$ (dynes/cm$^2$)  &   0.33   &   0.36   &   0.28   \\
\end{tabular}
\vspace{0.5cm}
\caption{\small Here the vacancy formation energy, latent heat 
values and elastic constants are presented for the original version 
of the EAM Al potential and the modified version.  $C_{11}$ and $C_{12}$ 
for the modified version (with the experimental melting temperature) 
are closer to their experimental values.  $C_{44}$ seems to get slightly 
worse. }  \label{table:c11}
\end{table}
\else
\end{multicols}
\fi


\begin{thebibliography}{10}
\bibitem{brian} Author to whom correspondence should be addressed.
\end{thebibliography}
\end{document}